\documentclass[a4paper,UKenglish,cleveref, autoref, thm-restate]{lipics-v2021}

\pdfoutput=1 
\hideLIPIcs  

\title{Distributive Laws of Monadic Containers} 

\author{Chris Purdy}{Royal Holloway University of London, UK}{christopher.purdy@rhul.ac.uk}{https://orcid.org/0009-0007-6105-1775}{}

\author{Stefania Damato}{School of Computer Science, University of Nottingham, UK}{stefania.damato@nottingham.ac.uk}{https://orcid.org/0009-0001-7182-5304}{}

\authorrunning{C. Purdy and S. Damato} 

\Copyright{Christopher Purdy and Stefania Damato. Authors' Accepted Manuscript} 



\acknowledgements{The authors would like to thank Thorsten Altenkirch, Dan Marsden, Sti\'ephen Pradal, and Reuben Rowe for helpful discussions on this work, Tom de Jong for inspiring our use of the \formalised\ symbol for formalised statements, and the anonymous reviewers for their suggestions.}

\nolinenumbers 

\EventEditors{Corina C\^{i}rstea and Alexander Knapp}
\EventNoEds{2}
\EventLongTitle{11th Conference on Algebra and Coalgebra in Computer Science (CALCO 2025)}
\EventShortTitle{CALCO 2025}
\EventAcronym{CALCO}
\EventYear{2025}
\EventDate{June 16--18, 2025}
\EventLocation{University of Strathclyde, UK}
\EventLogo{}
\SeriesVolume{342}
\ArticleNo{2}

\usepackage{packages}

\begin{document}

\maketitle

\begin{abstract}

Containers are used to carve out a class of strictly positive data types in terms of shapes and positions. They can be interpreted via a fully-faithful functor into endofunctors on $\cat{\Set}$. Monadic containers are those containers whose interpretation as a $\cat{\Set}$ functor carries a monad structure. The category of containers is closed under container composition and is a monoidal category, whereas monadic containers do not in general compose. 

In this paper, we develop a characterisation of distributive laws of monadic containers. Distributive laws were introduced as a sufficient condition for the composition of the underlying functors of two monads to also carry a monad structure. Our development parallels Ahman and Uustalu's characterisation of distributive laws of directed containers, i.e.\ containers whose $\cat{\Set}$ functor interpretation carries a \textit{comonad} structure. Furthermore, by combining our work with theirs, we construct characterisations of mixed distributive laws (i.e.\ of directed containers over monadic containers and vice versa), thereby completing the `zoo' of container characterisations of (co)monads and their distributive laws. We have found these characterisations amenable to development of existence and uniqueness proofs of distributive laws, particularly in the mechanised setting of Cubical Agda, in which most of the theory of this paper has been formalised.
\end{abstract}

\section{Introduction}

Containers, introduced by Abbott et al.\ \cite{abbott-containers}, give an algebraic presentation of a wide class of strictly positive data types. Often, reasoning about strictly positive data types in terms of their container representation is simpler than their functorial representation -- for example, transformations between container functors constructed from container morphisms are automatically natural.

Monads have received a lot of attention in functional programming \cite{haskell} and denotational semantics for their ability to model a wide range of programmatic side-effects \cite{moggi-monad}. In practice, it is rare for side-effects to appear individually -- developing a way of composing monads is useful for situations where multiple effects are interleaved. In general, however, the composition of two monads need not result in another monad. Distributive laws \cite{beck-composition} were developed as a sufficient condition for such a composition to form a monad, thereby ensuring that the corresponding side-effects are interleaved in a coherent way. Constructing distributive laws is known to be quite difficult due to the complexities involved in checking their axioms, besides the fact that some monads, even if they are composable, do not admit a distributive law in the first place \cite[Remark 4.19]{zwart-no-go-theorems}. As a result, various work has been done on different approaches for constructing distributive laws \cite{manes-mulry-comp-one, manes-mulry-comp-two}, and for identifying cases where there are none \cite{zwart-no-go-theorems, karamlou-shah-distr-laws}. 

In this paper, we develop a characterisation of distributive laws of monadic containers \cite{uustalu-monad-cont}, i.e.\ containers whose interpretation carries a monad structure, with the goal of providing an algebraic way of reasoning about distributive laws between strictly positive data types. We build on similar characterisations in the literature; Ahman, Chapman, and Uustalu develop directed containers \cite{ahman-directed-cont}, i.e.\ 
containers whose interpretation carries a \textit{comonad} structure, and the first and last authors provide a characterisation of their distributive laws \cite{ahman-distr-laws}. Uustalu also develops monadic containers \cite{uustalu-monad-cont}, but to our knowledge, no work has been done on characterising their distributive laws. Our work parallels the development in  \cite{ahman-distr-laws} for monadic containers and therefore closes this gap. Furthermore, by combining our work with \cite{ahman-distr-laws}, we construct characterisations of mixed distributive laws (i.e.\ of directed containers over monadic containers and vice versa), thereby completing the `zoo' of container characterisations of (co)monads and their distributive laws.

\subsection*{Formalisation}

Since our representation of distributive laws involves a list of highly dependent equalities, it lends itself well to formalisation in a proof assistant such as Cubical Agda \cite{cubical-agda}, in which we have formalised our characterisations as well as those in \cite{ahman-directed-cont,ahman-distr-laws}. The formalised statements are annotated by a \formalised\ symbol, which is a clickable link to the corresponding statement in the formalisation. The code can be found at \rurl{github.com/chrisjpurdy/distr-laws-of-monadic-containers}, and an HTML rendered version can be found at \rurl{stefaniatadama.com/distr-law-mnd-cont-html/}. 
While the aformentioned code implicitly assumes that we are dealing with h-sets throughout, we are currently working on a version of this code that makes this explicit and uses definitions from the Cubical library \cite{The_Agda_Community_Cubical_Agda_Library_2025}. This version can be found at \rurl{github.com/stefaniatadama/cubical/tree/distr-laws}.

\subsection*{Setting of our work \& Notation}

We work in cubical type theory \cite{CCHM, cubical-agda} but our work holds in any intensional Martin-L\"of type theory by assuming extensionality principles that are provable in cubical type theory, namely function extensionality (in particular, we do \textit{not} use univalence). Throughout the paper, we use several type-theoretic notations made explicit below. 

\begin{itemize}
	\item We use  $\sum_{a:A} B\, a$
	for dependent sums and $\prod_{a:A} B\,a$
	for dependent products, their non-dependent counterparts represented as $A\times B$ and $A\to B$ respectively. $A+B$ denotes the sum type of $A$ and $B$. The notation $(a , b)$ refers to an element of a product type, $\lambda x. f\, x$ an element of a function type, and $\inl\,a$ and $\inr\,b$ elements of a sum. $\pi_1$ and $\pi_2$ are the first and second projections out of a product type.
	
	\item The symbol $\coloneqq$ refers to definitional equality,
	$=$ refers to propositional equality, and $\cong$ refers to isomorphism of types.
	
	\item The empty type is denoted by $\bot$ and the unit type by $\top$ with element $\star$. $\Fin{n}$ is the type of finite sets of size $n$.
	
	\item The category whose objects are (small) sets and whose morphisms are functions is denoted by \cat{Set}, while $\mathsf{Set}$ denotes the universe of homotopy-sets (h-sets), or 0-types.
	
	\item The category of endofunctors on \cat{Set} is denoted by $[\cat{Set}, \cat{Set}]$.
	
	\item A few times we mention that an equality `holds up to' another equality, by which we mean that we have a heterogenous equality, or a path `lying over' another path. For more details on this, see \cite[Section 2.3]{hott-book}.
\end{itemize}

\subsection*{Organisation of the paper}

The paper is organised as follows. After mentioning some related work in \cref{sec:related-work}, we briefly review containers and monadic containers in \cref{sec:monadic-containers}. In \cref{sec:distributive-laws}, we characterise distributive laws of monadic containers and provide some examples using this characterisation, and in \cref{sec:composing} we look at compatible composites of monadic containers. In \cref{sec:mixed-laws}, we use an existing characterisation of distributive laws of directed containers to characterise mixed distributive laws. In  \cref{sec:no-go} we provide a starting point for using this characterisation to prove no-go theorems for monadic containers, before concluding in \cref{sec:conclusion}.

\section{Related work}\label{sec:related-work}

Containers are related to polynomial functors, as studied by Gambino and Hyland \cite{gambino-hyland-poly-functors}. Specifically, polynomial functors on the LCCC of $\mathsf{Set}$s are equivalent to indexed container functors, and in a special case we get (ordinary) container functors (when the object $I$ in \cite[Section 5]{gambino-hyland-poly-functors} is the terminal object). Polynomial monads (polynomial functors equipped with \emph{cartesian} monad maps) have been studied by Gambino and Kock \cite{gambino-kock-poly-monad}, but to our knowledge there is no characterisation of when polynomial monads can be composed via a distributive law or otherwise. Monadic containers fail to be a special case of polynomial monads, as monadic containers are not required to be cartesian. However, we do have that cartesian monadic containers, referred to as $\Sigma$-universes by Altenkirch and Pinyo \cite{altenkirch-monad-univ}, are special cases of polynomial monads. Awodey first mentions the connection between lax $\Sigma$-universes and a monad structure on polynomial functors in \cite[Remark 13]{awodey-natural-models}.

Manes and Mulry \cite{manes-mulry-comp-one,manes-mulry-comp-two} have developed general theorems concerning the existence of distributive laws. Zwart and Marsden \cite{zwart-no-go-theorems,zwart-thesis} have developed `no-go theorems' for monads that can be represented as algebraic theories, filling many holes in an extended version of the Boom hierarchy \cite{bunkenburg-boom}. Algebraic theories are disjoint from monadic containers in the sense that all monads representable by algebraic theories are finitary. On the other hand, the container $\cont{\top}{\lconst{A}}$ can be uniquely equipped with monadic container data, and the extension of this container is not finitary if $A$ is non-finite \cite{kock-notes}. Ahman, Chapman, and Uustalu's characterisation of directed containers \cite{ahman-directed-cont} has been used by Karamlou and Shah to develop no-go theorems relevant to finite model theory \cite{karamlou-shah-distr-laws}.

Directed containers and certain theorems in \cite{ahman-directed-cont} have already been formalised in vanilla Agda \cite{directed-containers-formalisation}, but to our knowledge their distributive laws have not. We formalise directed containers and their distributive laws alongside our own developments in Cubical Agda, but do not recreate all proofs included in \cite{directed-containers-formalisation}.

\section{Monadic containers}\label{sec:monadic-containers}

In this section, we briefly review containers and their functor representation, we state the definition of monadic containers we will be using, and provide some examples of both.

\subsection{Containers}

The purpose of containers \cite{abbott-containers, abbott-w, altenkirch-cats-of-conts, altenkirch-indexed-conts} is to provide a uniform way to represent strictly positive data types. By a `strictly positive type' we roughly mean a type $X$ whose constructors are only allowed to have $X$ appear in input types that are arrows if it appears on their right. Allowing non-strictly positive types in our theories can be problematic and potentially lead to contradictions \cite[Section\ 5.6]{hott-book}. In this sense, containers carve out a class of nicely behaved types, and they express the fact that any such type can be fully represented by a set of `shapes' $S$ and a family of `positions' $P$ over those shapes, at which data can be stored. 

\begin{definition}
	A \ul{container} is given by a type $S : \mathsf{Set}$ and a family of types $P : S \to \mathsf{Set}$, which we write as $\cont{S}{P}$.
\end{definition}

\begin{example}[\flink{ContainerExamples.html\#722}] 
	The container representation of the maybe type, which can either contain a value or indicate that there is no value, is the following. 
	\[\cont{\top + \top}{\lambda}
	 \begin{cases}
			\inl\,\star\to\bot\\ 
			\inr\,\star\to\top
		\end{cases}\]
	The shape $\top + \top$ represents a choice between having a value or not. If the shape is $\inl\,\star$, this represents the absence of a value, so the type of positions is $\bot$ as there is no more data to supply. If the shape is $\inr\,\star$, we do have a value, so the type of positions is $\top$ since there is one position for a singular piece of data.
\end{example}

\begin{example}[\flink{ContainerExamples.html\#782}] 
	The container representation of the list type is $\cont{\mathbb{N}}{\justFin}$. The shape of a list is a natural number $n$ representing its length, and there are $n$-many positions for data to be stored in a list of length $n$, represented by the set $\Fin{n}$.
\end{example}

Containers together with the following morphisms form a category \cat{Cont}.

\begin{definition}
	A \ul{container morphism} $(\cont{S}{P})\to(\cont{T}{Q})$ is a pair 
	\begin{align*}
		u &\colon S\to T\\
		f &\colon \nprod_{s:S} Q\,(u\,s)\to P\,s
	\end{align*}
	written as $\cont{u}{f}$. 
\end{definition}

\begin{example}
There is a container morphism $\cont{\mathbb{N}}{\justFin}\to \cont{(\top+\mathbb{N})}{\lambda x . \begin{cases}
    \inl\,\star \to \bot \\
    \inr\,n \to \Fin{n}
\end{cases}}$ that `takes the tail' of a list container, given by
\begin{align*}
    u\, 0 &\coloneqq \inl\,\star\\
    u\, (n + 1) &\coloneqq  \inr\,n \qquad\qquad\qquad\qquad f\, (n + 1)\,p \coloneqq p + 1
\end{align*}
\end{example}

The function $u$ describes the change to the length of the list (`failing' if the list is empty), and $f$ maps positions in the tail of the list to positions in the original list.

Every container has a functorial interpretation in the category $[\cat{Set}, \cat{Set}]$.

\begin{definition}\label{cont-interp}
	The \ul{container functor} associated to a container $\cont{S}{P}$ is the functor denoted by $\contfunc{S}{P}\colon \mathsf{Set} \to \mathsf{Set}$, with the following actions on objects and morphisms.
	\begin{itemize}
		\item Given an $X : \mathsf{Set}, \contfuncX{S}{P}{X}\coloneqq\underset{s:S}\nsum (P\, s\to X)$.
		\item Given $X, Y : \mathsf{Set}$ and a morphism $f\colon X\to Y$, for $s : S$ and $g\colon P\, s\to X$,
		\[\contfuncX{S}{P}{f}\,(s , g)\coloneqq (s , f\circ g).\]
	\end{itemize}
\end{definition}

The map $\llbracket \_ \rrbracket\colon \cat{Cont} \to[\cat{Set}, \cat{Set}]$ is extended to a fully-faithful functor by taking $\llbracket u \lhd f \rrbracket\,(s , g) := (u\,s , g \circ f)$ as the action on container morphisms.
Crucially for us, the category \cat{Cont} is monoidal, with the composition of containers as the monoidal product and $\cont{\top}{\lconst{\top}}$ as the unit.  

\begin{definition}\label{cont-composition}
Given containers $\cont{S}{P}$ and $\cont{T}{Q}$, their \ul{composite}, denoted $(\cont{S}{P}) \circ (\cont{T}{Q})$, is defined as the container
\[ 
    (\cont{S}{P}) \circ (\cont{T}{Q}) := \cont{\extcont{S}{P}\,T}{\Big (\lambda (s , f) . \nsum_{p : P\,s} Q\,(f\, p)\Big )} 
\]
\end{definition}

The functor $\llbracket \_ \rrbracket$ is monoidal, i.e.\ it preserves monoidal multiplication and unit.

\subsection{Monadic containers}

Monadic containers were developed by Uustalu \cite{uustalu-monad-cont} as a characterisation of monads whose underlying functors are container functors. We adopt conventions used by Altenkirch and Pinyo in \cite{altenkirch-monad-univ}, referring to them as monadic containers (rather than mnd-containers) and using their presentation as lax $\Sigma$-universes.

\begin{definition}[\flink{MndContainer.html\#1941}] \label{mcont} 
Let $\cont{S}{P}$ be a container.
A \ul{monadic container} on $\cont{S}{P}$ is a tuple $(\cont{S}{P}, \iota, \sigma, \pr)$ where
\begin{align*}
    &\iota : S\\
    &\sigma : \nprod_{s : S} (P\, s \to S) \to S\\
    &\pr : \nprod_{\imparg{s : S}} \nprod_{\imparg{f : P\, s \to S}} P\, (\sigma\, s\, f) \to \nsum_{p : P\, s}\, P\, (f\, p)
\end{align*}
satisfying the following equations:
{\small
\begin{align*}
    \sigma\, \iota\, (\lconst{s}) &= s 									& \sigma\, s\, (\lconst{\iota}) &= s\\
    \pr_2\, \imparg{\iota}\, \imparg{\lconst{s}}\, p &= p  &  \pr_1\, \imparg{s}\, \imparg{\lconst{\iota}}\, p &= p\\
\end{align*}\vspace{-30pt}
\begin{align*}
&\sigma\, s\, (\lambda p .\, \sigma\, (f\, p)\, (g \circ (p , -))) = \sigma\, (\sigma\, s\, f)\, (g \circ \pr) \\
& \pr_1\, \imparg{s}\, \imparg{\lambda p . \sigma\, (f\, p)\, (g \circ (p , -))}\, p = \pr_1\, \imparg{s}\, \imparg{f}\, (\pr_1\, \imparg{\sigma\, s\, f}\, \imparg{g \circ \pr}\, p) \\
&\pr_1\, \imparg{f\, (\pr_1\,p)}\, \imparg{\lambda p' . g\, (\pr_1\,p , p')}\, (\pr_2\, \imparg{s}\, \imparg{\lambda p . \sigma\, (f\, p)\, (g \circ (p , -))}\, p) =\\
&\hspace{7.9cm} \pr_2\, \imparg{s}\, \imparg{f}\, (\pr_1\, \imparg{\sigma\, s\, f}\, \imparg{g \circ \pr}\, p)\\
&\pr_2\, \imparg{f\,(\pr_1\,p)}\, \imparg{\lambda p' . g\, (\pr_1\,p , p')}\, (\pr_2\, \imparg{s}\, \imparg{\lambda p . \sigma\, (f\, p)\, (g \circ (p , -))}\, p) = \pr_2\, \imparg{\sigma\, s\, f}\, \imparg{g \circ \pr}\, p
\end{align*}
}
where we use the shorthands $\pr_i := \pi_i \circ \pr$, and $(p , -) := \lambda p' . (p , p')$. The equalities for $\pr$ hold up to the corresponding equalities for $\sigma$ above them.
\end{definition}

For clarity we include all implicit arguments in grey in the equations above, but we will often omit them when they can be inferred from the context.
For example, given $p : P\, (\sigma\, (\sigma\, s\, f)\, (g \circ \pr))$, we can write the last three equalities for $\pr$ as
\begin{align*}
    \pr_1\, p &= \pr_1\, (\pr_1\, p) & \pr_1\, (\pr_2\, p) &= \pr_2\, (\pr_1\, p) & \pr_2\, (\pr_2\, p) &= \pr_2\, p
\end{align*}

Abusing notation, we will refer to a monadic container $(\cont{S}{P}, \iota, \sigma, \pr)$ by its container $\cont{S}{P}$, when $\iota, \sigma$ and $\pr$ are clear from the context.

Monadic containers can be thought of as containers whose sets of shapes are pointed, and closed under taking `container-fulls' of shapes, in the sense that any element $(s , f) : \contfunc{S}{P}\,S$ gives a new shape $\sigma\, s\, f : S$. The role of $\pr$ is to specify how positions $p : P\,(\sigma\,s\,f)$ map to positions $p_1 : P\, s$ and $p_2 : P\,(f\,p_1)$.

Every monadic container can be interpreted as a monad on $\cat{\Set}$, this interpretation being an extension of the usual interpretation of containers as functors.

\begin{definition}\label{mcont-interp}
The \ul{monad interpretation} of a monadic container $(\cont{S}{P}, \iota, \sigma, \pr)$ is defined as the monad $(\contfunc{S}{P}, \eta, \mu)$ on \cat{Set}, denoted by $\contfunc{S}{P, \iota, \sigma, \pr }^\text{mc}$, or simply $\contfunc{S}{P}^\text{mc}$ when the monadic container structure is clear from the context, where
\begin{align*}
    &\eta : \Id \Rightarrow \contfunc{S}{P} & & \mu : \contfunc{S}{P} \circ \contfunc{S}{P} \Rightarrow \contfunc{S}{P}\\
    &\eta_A\, a := (\iota , \lconst{a}) & & \mu_A\, (s , f) := (\sigma\, s\, (\pi_1 \circ f) , \lambda p . \pi_2\,(f\,(\pr_1\, p))\,(\pr_2\, p))
\end{align*}  
\end{definition}

Monadic containers can also be seen as lax versions of (Tarski-style) type universes closed under singleton and dependent sum types \cite{altenkirch-monad-univ}. Shapes are interpreted as codes for types, and the position family is the map that interprets each code as a concrete type. $\iota$ is the code for the singleton type, and $\sigma$ constructs codes for $\Sigma$-types.

\begin{definition}
A \ul{$\Sigma$-universe} $(\cont{S}{P}, \iota, \un, \sigma, \pr)$ is given by a monadic container $(\cont{S}{P}, \iota, \sigma, \pr)$ with the isomorphisms
\begin{align*}
    &\un : P\, \iota \cong \top\\
    &\pr : \nprod_{\imparg{s : S}} \nprod_{\imparg{f : P\, s \to S}} P\, (\sigma\, s\, f) \cong \nsum_{p : P\, s}\ P\, (f\, p)
\end{align*}
i.e.\ where for all $s$ and $f$, $\pr\, \imparg{s}\, \imparg{f}$ is an isomorphism, and $P\, \iota$ is isomorphic to $\top$. 
\end{definition}

The container examples we describe at the start of this section can each be equipped with monadic container structure.

\begin{example} \label{list-mcont}
The list container 
$\cont{\nat}{\Fin{}}$\hspace{-2.5pt}
can be extended to a monadic container by taking
\begin{align*}
    \iota &:= 1\\
    \sigma\, n\, f &:= f\, 0 + \dots + f\, (n-1)\\ 
    \pr_1\,\imparg{n}\,\imparg{f}\, p &:= \max\, \{ i \in [0..n)\ |\ f\, 0 + \dots + f\, (i - 1) \le p \}\\  
    \pr_2\,\imparg{n}\,\imparg{f}\, p &:= p - (f\, 0 + \dots + f\, ((\pr_1\,\imparg{n}\,\imparg{f}\, p) - 1)). 
\end{align*}
This is also an example of a $\Sigma$-universe. The monad interpretation of this container is isomorphic to the list monad with concatenation.
\end{example}

\begin{example}[\flink{ContainerExamples.html\#2768}] 
Given some set $S$, the container
$\cont{(S \to S)}{\lconst{S}}$
can be extended to a monadic container by taking
\begin{align*}
    \iota &:= \lambda x . x\\
    \sigma\,f\,g &:= \lambda x . g\,x\,(f\,x)\\
    \pr\,\imparg{f}\,x &:= ( x , f\,x ).
\end{align*}
The monad interpretation of this is the well-known state monad from functional programming.
\end{example}

\begin{example}[\flink{ContainerExamples.html\#4141}] \label{coprod-mcont} 
The container
$\cont{(\top + E)}{\Tr}$ of coproducts with $E$,
where 
\begin{align*}
	\Tr\, (\inl\,\star) &\coloneqq \top\\
	\Tr\, (\inr\,e) &\coloneqq \bot
\end{align*}
can be extended to a monadic container by taking
\begin{align*}
    \iota &:= \inl\,\star\\
    \sigma\,(\inl\, \star)\,f &\coloneqq f\,\star\\
    \sigma\,(\inr\, e)\, \_ &\coloneqq \inr\, e\\
    \pr\,\imparg{\inl\, \star}\,\star &:= (\star, \star).
\end{align*}
This is another example of a $\Sigma$-universe. The monad interpretation of this is known to the functional programming community as the exception monad.
\end{example}

\begin{example}[\flink{ContainerExamples.html\#1981}] 
Given a monoid $(A , \otimes, e)$, the container
$\cont{A}{\lconst{\top}}$
can be extended to a monadic container by taking
\begin{align*}
    \iota &:= e\\
    \sigma\,a\,f &:= a \otimes f\,\star\\
    \pr\,\star &:= (\star, \star).
\end{align*}
The monadic container equalities hold as a consequence of the monoid equalities. We call this the \ul{\textit{writer monadic container}}. Monadic containers on $\cont{A}{\lconst{\top}}$ are in bijection with monoids on $A$. The monad interpretation of this is typically called the writer monad.
\end{example}

We recall the definition of directed containers \cite{ahman-directed-cont}, for use in later sections of the paper.

\begin{definition}[\flink{DirectedContainer.html\#207}] \label{dcont}
Let $\cont{S}{P}$ be a container. A \underline{directed container} on $\cont{S}{P}$ is a tuple $(\cont{S}{P}, o , \oplus, \downarrow)$ where
\begin{gather}
\begin{align*}
o &: \nprod_{\imparg{s : S}} P\,s\\
{\downarrow} &: \nprod_{s : S} P\,s \to S\\
\oplus &: \nprod_{\imparg{s : S}} \nprod_{p : P\,s} P\,(s \downarrow p) \to P\,s
\end{align*}
\qquad\qquad
{ \small
\begin{align*}
    s \downarrow o &= s\\
    s \downarrow (p \oplus p') &= (s \downarrow p) \downarrow p'\\
    p \oplus o &= p\\
    o \oplus p &= p\\
    (p \oplus p') \oplus p'' &= p \oplus (p' \oplus p'').
\end{align*}
}
\end{gather}
\end{definition}

\section{Distributive laws}\label{sec:distributive-laws}

The question of whether two monads compose has applications in many areas. In functional programming and denotational semantics of effectful languages, for example, one might consider composition of two (or more) notions of side effect. As remarked in \cite[Remark 1.1]{zwart-no-go-theorems}, distributive laws provide a particularly nice way to compose monads, that imbue the resulting composite monad with a variety of desirable properties.

Distributive laws were first introduced by Beck \cite{beck-composition} as a sufficient condition for the composition of the underlying functors of two monads to carry a monad structure.

\begin{definition}\label{cat-distr-law}
Let $\mathbf{S} = (S , \eta^S, \mu^S)$ and $\mathbf{T} = (T , \eta^T, \mu^T)$ be monads.
A \underline{distributive law} of $\mathbf{S}$ over $\mathbf{T}$ is a natural transformation $\gamma : TS \Rightarrow ST$ such that the following diagrams commute.
\[
\begin{tikzcd}[row sep=25pt]
	 & S\arrow[swap]{dl}{\eta^T S}\arrow{dr}{S \eta^T} & & & & & T\arrow[swap]{dl}{T\eta^S}\arrow{dr}{\eta^S T} &\\
	 TS\arrow[swap]{rr}{\gamma} & & ST & & & TS\arrow[swap]{rr}{\gamma} & & ST\\[10pt]
	 TTS\arrow[swap]{d}{\mu^T S}\arrow{r}{T\gamma} & TST\arrow{r}{\gamma T} & STT\arrow{d}{S\mu^T} & & & TSS\arrow{r}{\gamma S}\arrow[swap]{d}{T\mu^S} & STS\arrow{r}{S\gamma} & SST\arrow{d}{\mu^S T}\\
	 TS\arrow[swap]{rr}{\gamma} & & ST & & & TS \arrow[swap]{rr}{\gamma}& & ST
\end{tikzcd} 
\]
\end{definition}

In our approach, we specialise \cref{cat-distr-law} to the case when $S$ and $T$ are container functors, making $\mathbf{S}$ and $\mathbf{T}$ monads on container functors.  We rely on the fact that both the container interpretation functor $\llbracket \_ \rrbracket\colon \cat{Cont}\to [\cat{Set}, \cat{Set}]$ and the monadic container interpretation functor $\llbracket \_ \rrbracket^\text{mc}\colon \cat{MCont} \to \cat{Monad} (\cat{Set})$ are fully-faithful \cite{abbott-containers, uustalu-monad-cont}, as well as monoidality of \cat{Cont} and $\llbracket \_ \rrbracket$. This lets us directly interpret the diagrams in \cref{cat-distr-law} as diagrams in \cat{Cont},
and gives us that each monadic container distributive law of $\cont{S}{P}$ over $\cont{T}{Q}$ corresponds to a unique monad distributive law of $\contfunc{S}{P}^\text{mc}$ over $\contfunc{T}{Q}^\text{mc}$.

\begin{definition}[\flink{MndDistributiveLaw.html\#298}] \label{m-cont-distr-law} 
Let $(\cont{S}{P}, \iota^S, \sigma^S, \pr^S)$ and $(\cont{T}{Q}, \iota^T, \sigma^T, \pr^T)$ be monadic containers.
A \ul{monadic container distributive law} of $\cont{T}{Q}$ over\footnote{The ordering of the monadic containers in this terminology follows that used by Beck.} $\cont{S}{P}$ is given by the data
\begin{align*}
    &u_1 : \nprod_{s : S} (P\, s \to T) \to T \\
    &u_2 : \nprod_{s : S} \nprod_{f : P\, s \to T} Q\, (u_1\, s\, f) \to S \\
    &v_1 : \nprod_{\imparg{s : S}} \nprod_{\imparg{f : P\, s \to T}} \nprod_{q : Q\, (u_1\, s\, f)} P\, (u_2\, s\, f\, q) \to P\, s \\
    &v_2 : \nprod_{\imparg{s : S}} \nprod_{\imparg{f : P\, s \to T}} \nprod_{q : Q\, (u_1\, s\, f)} \nprod_{p : P\, (u_2\, s\, f\, q)} Q\, (f\, (v_1\, q\, p))
\end{align*} 
which satisfy the following equalities.
{\small
\begin{align*}
    u_1\, \iota^S\, (\lconst{t}) &= t \tag{unit-$\iota S$-$s_1$}\\
    u_2\, \iota^S\, (\lconst{t}) &=\lconst{\iota^S} \tag{unit-$\iota S$-$s_2$}\\
    v_1\, \imparg{\iota^S}\, \imparg{\lambda \_. t}\, q\, p &= p \tag{unit-$\iota S$-$p_1$}\\
    v_2\, \imparg{\iota^S}\, \imparg{\lambda \_. t}\, q\, p &= q \tag{unit-$\iota S$-$p_2$}\\
    \\
    u_1\, s\, (\lambda \_. \iota^T) &= \iota^T \tag{unit-$\iota T$-$s_1$}\\
    u_2\, s\, (\lambda \_. \iota^T) &= \lconst{s} \tag{unit-$\iota T$-$s_2$}\\
    v_1\, \imparg{s}\, \imparg{\lambda \_. \iota^T}\, q\, p &= p \tag{unit-$\iota T$-$p_1$}\\
    v_2\, \imparg{s}\, \imparg{\lambda \_. \iota^T}\, q\, p &= q \tag{unit-$\iota T$-$p_2$}\\
    \\
    u_1\, (\sigma^S\, s\, f)\, (g \circ \pr^S) &= u_1\, s\, (\lambda p . u_1\, (f\, p)\, (g \circ (p,-))) \tag{mul-$S$-$s_1$}\\
    u_2\, (\sigma^S\, s\, f)\, (g \circ \pr^S)\, q &= \sigma^S\,(u_2\, s\, (\lambda p .u_1\, (f\, p)\, (g \circ (p,-)))\, q)\\ 
    &\qquad\;\;\, (\lambda p . u_2\, (f\, (v_1\, q\, p))\, (g \circ (v_1\, q\, p , -))\, (v_2\, q\, p)) \tag{mul-$S$-$s_2$}\\  
    \pr^S_1\, (v_1\, q\, p) &= v_1\, q\, (\pr^S_1\, p) \tag{mul-$S$-$p_1$}\\
    \pr^S_2\, (v_1\, q\, p) &= v_1\, (v_2\, q\, (\pr^S_1\, p))\, (\pr^S_2\, p) \tag{mul-$S$-$p_{21}$}\\
    v_2\, q\, p &= v_2\, (v_2\, q\, (\pr^S_1\, p))\, (\pr^S_2\, p) \tag{mul-$S$-$p_{22}$}\\
    \\
    u_1\, s\, (\lambda p . \sigma^T\, (f\, p)\, (g \circ (p,-))) &= \sigma^T\, (u_1\, s\, f)\, (\lambda q . u_1\, (u_2\, s\, f\, q)\, (g \circ v\, q)) \tag{mul-$T$-$s_1$}\\
    u_2\, s\, (\lambda p . \sigma^T\, (f\, p)\, (g \circ (p,-)))\, q &= u_2\, (u_2\, s\, f\, (\pr^T_1\, q))\, (g \circ v\, q)\, (\pr^T_2\, q) \tag{mul-$T$-$s_2$}\\
    v_1\, (\pr^T_1\, q)\, (v_1\, (\pr^T_2\, q)\, p) &= v_1\, q\, p \tag{mul-$T$-$p_1$}\\
    v_2\, (\pr^T_1\, q)\, (v_1\, (\pr^T_2\, q)\, p) &= \pr^T_1\, (v_2\, q\, p) \tag{mul-$T$-$p_{21}$}\\
    v_2\, (\pr^T_2\, q)\, p &= \pr^T_2\, (v_2\, q\, p) \tag{mul-$T$-$p_{22}$}
\end{align*}
}
\end{definition}

We will use the shorthands 
\begin{align*}
u\,s\,f &\coloneqq (u_1\, s\, f , u_2\, s\, f) \\
v\, \imparg{s}\, \imparg{f}\,q\,p &\coloneqq (v_1\, \imparg{s}\, \imparg{f}\, q\, p , v_2\, \imparg{s}\, \imparg{f}\, q\, p),
\end{align*}
primarily to simplify stating examples of distributive laws.

We refer to equalities whose names end in $s_1$ or $s_2$ collectively as `shape' equalities. The ones ending with names in $p_1$, $p_2$, $p_{21}$, or $p_{22}$ are referred to collectively as `position' equalities.

These may seem rather unwieldy, but often in applications of this characterisation many of the equations simplify considerably. For example, when constructing a distributive law, if we define a $u_1, u_2$ such that we can take $v_1\, \imparg{s}\, \imparg{f}\, q\, p = p$ and $v_2\, \imparg{s}\, \imparg{f}\, q\, p = q$, then all the position equalities hold definitionally, and only the shape equalities must be proven (like in \cref{coprod-distr-law}). The distributive law uniqueness lemmas (\cref{unique-coprod-distr-law,reader-distr-unique,writer-reader-mixed-law-unique}) and no-go theorem in Section 7 only require consideration of a few of these equalities.

While \cref{cat-distr-law} is a concise way of stating what a distributive law is, expanding this definition to check the naturality of $\gamma$ and commutativity of the 4 diagrams turns out to be cumbersome in practice. Further, our phrasing of distributive laws lends itself to formalisation in a proof assistant. In Cubical Agda, the dependencies between equations can be explicitly stated in terms of dependent paths, and we can prove properties of distributive laws without descending into ``transport hell''. 

\begin{example}[\flink{DistributiveLawExamples.html\#728}] \label{coprod-distr-law} 
There is a monadic container distributive law of $\cont{S}{P}$ over $\cont{(\top + E)}{\Tr}$, for any $E$, $S$, and $P$, given by
\begin{align*}
    u\,(\inl\,\star)\,f &\coloneqq (f\, \star, \lconst{\inl\,\star})\\
    u\,(\inr\, e)\,\_ &\coloneqq (\iota^S, \lconst{\inr\, e})\\
    v\,\imparg{\inl\,\star}\,\imparg{f}\,p \,\star &\coloneqq (\star , p).
\end{align*} 
\end{example}

In fact, this is the only monadic container distributive law of this type, and the characterisation in \cref{m-cont-distr-law} offers a succinct proof of this:

\begin{lemma}[\flink{DistributiveLawExamples.html\#3242}] \label{unique-coprod-distr-law} 
The distributive law in \cref{coprod-distr-law} is the unique one of $\cont{S}{P}$ over $\cont{(\top + E)}{\Tr}$ for any $E$, $S$, and $P$.
\begin{proof}
    We first note that $(\top \to S) \cong S$, and that $\bot \to S$ is contractible ($(\bot \to S) \cong \top$, since it has a point $\lconst{\iota^S}$). Using these facts and the equalities unit-$\iota S$-$s_1$, unit-$\iota S$-$s_2$, unit-$\iota T$-$s_1$, and unit-$\iota T$-$s_2$, $u$ is uniquely specified up to function extensionality. We can use the same reasoning and laws unit-$\iota S$-$p_1$ and unit-$\iota S$-$p_2$ to see that $v$ is also uniquely specified up to function extensionality.
\end{proof}
\end{lemma}

\begin{example}[\flink{DistributiveLawExamples.html\#9150}] \label{reader-distr} 
There is a monadic container distributive law of $\cont{\top}{\lconst{A}}$ over $\cont{S}{P}$ for any $A$, $S$, and $P$ given by
\begin{align*}
    u\,s\,f &\coloneqq (\star, \lconst{s})\\
    v\,a\,p &\coloneqq (p , a).
\end{align*}
\end{example}

\begin{lemma}[\flink{DistributiveLawExamples.html\#10377}] \label{reader-distr-unique} 
The distributive law in \cref{reader-distr} is the unique one of $\cont{\top}{\lconst{A}}$ over $\cont{S}{P}$ for any $A$, $S$, and $P$.
\begin{proof}
Follows from the isomorphisms $\top \cong (P\,s \to \top)$ for any $s : S$, and the unit-$\iota T$-$\dots$ equalities for monadic container distributive laws.
\end{proof}
\end{lemma}

Ahman and Uustalu in \cite{ahman-distr-laws} noticed that distributive laws of directed containers generalise \emph{matching pairs of monoid actions}, and the composition of directed containers via a distributive law generalises the \emph{Zappa-Sz\'ep product} of monoids \cite{brin-zappa-szep}. The following example shows how distributive laws and composition of monadic containers generalise the same constructions, but in a slightly different way, which is unsurprising but nice to see. 

\begin{example}\label{writer-monoid-d-law}
Given monoids $(A, \otimes^A, \iota^A)$ and $(B, \otimes^B, \iota^B)$, and a matching pair of monoid actions $(\alpha : A \times B \to A, \beta : A \times B \to B)$, there is a distributive law of $\cont{B}{\lconst{\top}}$ over $\cont{A}{\lconst{\top}}$ given by
\begin{align*}
    u\,a\,b &\coloneqq (\beta\,(a , b\, \star), \lconst{\alpha\,(a , b\, \star)})\\
    v\,\star\,\star &\coloneqq (\star , \star)
\end{align*}
where the distributive law equalities follow from the equalities for the matching pair of monoid actions (up to the isomorphism $(\top \to B) \cong B$). Conversely, any distributive law of this type specifies a matching pair of monoid actions for the relevant monoids. 
\end{example}

\section{Composing with distributive laws}\label{sec:composing}

Along with distributive laws, Beck also introduced the equivalent notion of compatible composites \cite{beck-composition}, as a way to specify when a \emph{given} monad can be considered a composite of two others. In this section, we characterise these in terms of monadic containers. Often, we will use the shorthand $\contdia{P}{Q} := \lambda (s , f) . \nsum_{p : P\,s} Q\,(f\,p)$ to simplify the presentation of nested $\Sigma$-types.

\begin{definition}[\flink{MndCompatibleComposite.html\#344}] 
A \ul{compatible composite monadic container} of $(\cont{S}{P}, \iota^S, \sigma^S, \pr^S)$ over $(\cont{T}{Q}, \iota^T, \sigma^T, \pr^T)$ is a pair of maps
\begin{align*}
    &\sigma : \nprod_{x : \extcont{S}{P}{T}} (\contdia{P}{Q}\,x \to \extcont{S}{P}\, T) \to \extcont{S}{P}\, T\\
    &\pr : \nprod_{\imparg{x : \extcont{S}{P}{T}}} \nprod_{\imparg{f : \contdia{P}{Q}\,x \to \extcont{S}{P}\, T}} \contdia{P}{Q}\,(\sigma\, x\, f) \to \nsum_{p : \contdia{P}{Q}\,x} \contdia{P}{Q}\,(f\,p)
\end{align*}
where: 
\begin{itemize}
    \item $((\cont{S}{P}) \circ (\cont{T}{Q}), (\iota^S , \lconst{\iota^T}), \sigma, \pr)$ is a monadic container
    \item the container morphisms $\lambda s . (s , \lconst{\iota^T}) \lhd \pi_1$ and $\lambda t. (\iota^S , \lconst{t}) \lhd \pi_2$ are monadic container morphisms, as defined in \cite{uustalu-monad-cont}
    \item the middle unitary laws hold: \begin{align*}
        (s , f) &= \sigma\,(s , \lconst{\iota^T})\,(\lambda p . (\iota^S , \lconst{f\,(\pi_1\, p)}))\\
        (q , p) &= (\pi_1\, (\pr_1\,(q, p)) , \pi_2\, (\pr_2\,(q , p)))
    \end{align*} where the second equality is dependent on the first.
\end{itemize}
\end{definition}

Distributive laws of monadic containers and their interpreting monads are in bijection due to  $\llbracket\_\rrbracket\colon\cat{Cont}\to[\cat{Set},\cat{Set}]$ being full and faithful. Compatible composite monadic containers and their interpreting monads are also in bijection, since $\cont{\sigma}{\pr}$ is a container morphism, the collection of which is in bijection with natural transformations between container functors, again by $\llbracket\_\rrbracket$ being full and faithful. Therefore, Beck's result that in the monad setting distributive laws and compatible composites are equivalent also immediately implies that they are equivalent in the monadic container setting. However, we would still like to give a direct construction of the compatible composite monadic container we obtain from a monadic container distributive law, and vice versa.

\begin{proposition}[Compatible composite from a distributive law, partially formalised \flink{MndDistrLawToCompatibleComposite.html\#8951}]\label{distr-law-to-composite}
Given monadic containers  $(\cont{S}{P}, \iota^S, \sigma^S, \pr^S)$ and $(\cont{T}{Q}, \iota^T, \sigma^T, \pr^T)$, and a monadic container distributive law $(u_1 , u_2 , v_1 , v_2)$ of $\cont{S}{P}$ over $\cont{T}{Q}$, we have a compatible composite monadic container given by
\begin{alignat*}{3}
    \sigma\, (s , f)\, g &\coloneqq\ & (&\sigma^S\, s\, (\lambda p . u_1\, (f\, p)\, (g_1 \circ (p,-))) ,\\
        &&&\lambda p . \sigma^T\, 
        (u_2\, (f\, (\pr^S_1\, p))\, (g_1 \circ (\pr^S_1\, p, -))\, (\pr^S_2\, p))\\ 
        &&&\quad\;(\lambda q . g_2\, (\pr^S_1\, p , v_1\, (\pr^S_2\, p)\, q)\, (v_2\, (\pr^S_2\, p)\, q)))\\
    \pr\, (p, q) &\coloneqq\ & (&(\pr^S_1\, p , v_1\, (\pr^S_2\, p)\ (\pr^T_1\, q)),\\ 
    &&&(v_2\, (\pr^S_2\, p)\, (\pr^T_1\, q) , \pr^T_2\, q))
\end{alignat*}
where $g_i \coloneqq \pi_i \circ g$.
\begin{proof}
	Derivations of the `associativity' equalities are in \cref{appendixB}. Proofs of the remaining equalities are included in the Cubical Agda formalisation. 
\end{proof}
\end{proposition}

\begin{proposition}[Distributive law from a compatible composite]
Given monadic containers  $(\cont{S}{P}, \iota^S, \sigma^S, \pr^S)$ and $(\cont{T}{Q}, \iota^T, \sigma^T, \pr^T)$, and a compatible composite monadic container $(\sigma, \pr)$ of $\cont{S}{P}$ over $\cont{T}{Q}$, then we have a monadic container distributive law given by
\begin{align*}
	u\, s\, f &\coloneqq \sigma\, (\iota^T, \lconst{s})\, (\lambda\, p. (f\, (\pi_2\, p), \lconst{\iota^S}))\\
	v\, q\, p &\coloneqq (\pi_2\, (\pr_1\, (q, p)) , \pi_1\, (\pr_2\, (q, p))).
\end{align*}
\end{proposition}

To see that \cref{distr-law-to-composite} generalises the Zappa-Sz\'ep product of monoids, we consider the case in \cref{writer-monoid-d-law} of a distributive law between two writer monadic containers $\cont{A}{\lconst{\top}}$ and $\cont{B}{\lconst{\top}}$. The resulting compatible composite monadic container (constructed using \cref{distr-law-to-composite}) is again a writer monadic container $\cont{(A\times B)}{\lconst{\top}}$, whose corresponding monoid is a Zappa-Sz\'ep product of the monoids corresponding to $\cont{A}{\lconst{\top}}$ and $\cont{B}{\lconst{\top}}$.

\begin{example}
Given a set $A$ and monadic container $(\cont{S}{P}, \iota^S, \sigma^S, \pr^S)$, the compatible composite monadic container of $\cont{\top}{\lconst{A}}$ over $\cont{S}{P}$ arising from the distributive law in \cref{reader-distr} is given by
\begin{align*}
    \sigma\,(\star , f)\,g &\coloneqq (\star , \lambda a . \sigma^S\,(f\, a)\, (\lambda p . \pi_2\, (g\,(a , p)\,a)))\\
    \pr\,(a , p) &\coloneqq ((a , \pr^S_1\, p),(a , \pr^S_2\,p)).
\end{align*}
\end{example}

On a different note, composite monadic containers can be useful for considering extensions of $(1 ,\Sigma)$ type universes (modelled by $\Sigma$-universes \cite{altenkirch-monad-univ}). As an example, we consider how to augment a $\Sigma$-universe $(\cont{\mathcal{U}}{\text{El}}, \iota^{\mathcal{U}}, \un^{\mathcal{U}}, \sigma^{\mathcal{U}}, \pr^{\mathcal{U}})$ with codes for \emph{refinement types} \cite{jhala-refinement-types}. This can be done by considering compatible composites of $\cont{\mathcal{U}}{\text{El}}$ over the `Maybe' $\Sigma$-universe (\cref{coprod-mcont} where $E = \top$).

\begin{example}\label{predicate-univ}
Given a $\Sigma$-universe $(\cont{\mathcal{U}}{\text{El}}, \iota^{\mathcal{U}}, \un^{\mathcal{U}}, \sigma^{\mathcal{U}}, \pr^{\mathcal{U}})$, the composite monadic container corresponding to \cref{coprod-distr-law} when $E = \top$ is
{\small
\begin{align*}
    \hspace{-7pt}\iota &\coloneqq (\iota^{\mathcal{U}} , \lconst{\true})\\
    \hspace{-7pt}\sigma\,(s , f)\,g &\coloneqq \Big (
        \sigma^{\mathcal{U}}\,s\,\Big (\lambda p . \begin{cases}
            g_1\,(p , \star) & \text{if } f\,p = \true\\
            \iota^{\mathcal{U}} & \text{otherwise}
        \end{cases}\Big ) , 
        \lambda p . \begin{cases}
            g_2\,(\pr_1^{\mathcal{U}}\,p , \star)\,(\pr_2^{\mathcal{U}}\,p) & \text{if } f\,(\pr_1^{\mathcal{U}}\,p) = \true\\
            \false & \text{otherwise}
        \end{cases} 
    \Big )\\
    \hspace{-7pt}\pr\,(p , \star) &\coloneqq ((\pr^{\mathcal{U}}_1\,p , \star),(\pr^{\mathcal{U}}_2\, p ,\star))
\end{align*}
}
where $\true := \inl\,\star$ and $\false := \inr\,\star$. This composite monadic container is also a $\Sigma$-universe, as $\pr$ is an isomorphism (since $\pr^{\mathcal{U}}$ is an iso) and we have that $\contdia{\text{El}}{\Tr}\, \iota \cong \top$ (since $\un^{\mathcal{U}}$ is an iso).
\end{example}
To see how shapes in the above composite encode refinement types, consider a shape $(s , f) : \nsum_{s : \mathcal{U}} \El{s} \to (\top + \top)$. The shape $s$ is a code for the type $\El{s}$, and we can see $f$ as a \emph{predicate} on elements of $\El{s}$. The type that $(s , f)$ encodes is $\contdia{\text{El}}{\Tr}\,(s , f) \coloneqq \nsum_{x : \El{s}} \Tr\,(f\,x)$, whose elements are essentially the elements of $\El{s}$ for which the predicate $f$ holds.

This is an `augmentation' in the sense that all types encoded by $\mathcal{U}$ have unique codes in the composite universe. Given a shape $s : \mathcal{U}$, the composite shape $(s , \lconst{\true})$ codes for the type with elements from $\El{s}$ that satisfy the `always true' predicate -- this is clearly isomorphic to the type $\El{s}$.

\section{Mixed distributive laws}\label{sec:mixed-laws}

By combining our characterisation of monadic container distributive laws with Ahman and Uustalu's directed container distributive laws \cite[Section 4]{ahman-distr-laws} (\flink{DirectedDistributiveLaw.html\#246}), it is straightforward to obtain a characterisation of mixed container distributive laws.

\begin{definition}[\flink{MndDirectedDistributiveLaw.html\#360}] \label{mc-cont-distr-law}
Let $(\cont{T}{Q}, \iota, \sigma, \pr)$ be a monadic container and $(\cont{S}{P}, o , \oplus, \downarrow)$ be a directed container. We define a \ul{monadic-directed container distributive law} of $\cont{T}{Q}$ over $\cont{S}{P}$ as the data
\begin{align*}
    &u_1 : \nprod_{s : S} (P\, s \to T) \to T \\
    &u_2 : \nprod_{s : S} \nprod_{f : P\, s \to T} Q\, (u_1\, s\, f) \to S \\
    &v_1 : \nprod_{\imparg{s : S}} \nprod_{\imparg{f : P\, s \to T}} \nprod_{q : Q\, (u_1\, s\, f)} P\, (u_2\, s\, f\, q) \to P\, s \\
    &v_2 : \nprod_{\imparg{s : S}} \nprod_{\imparg{f : P\, s \to T}} \nprod_{q : Q\, (u_1\, s\, f)} \nprod_{p : P\, (u_2\, s\, f\, q)} Q\, (f\, (v_1\, q\, p))
\end{align*}
which satisfy the following equalities.
\begin{align*}
    u_1\, s\, f &= f\,(o\, \imparg{s}) \tag{unit-$oS$-$s$}\\
    v_1\, \imparg{s}\, \imparg{f}\, q\, (o\, \imparg{u_2\, s\, f\, q}) &= o\,\imparg{s} \tag{unit-$oS$-$p_1$}\\
    v_2\, \imparg{s}\, \imparg{f}\, q\, (o\, \imparg{u_2\, s\, f\, q}) &= q \tag{unit-$oS$-$p_2$}\\
    u_2\,s\,f\,q \downarrow p &= u_2\,(s \downarrow v_1\,q\,p)\, (\lambda p' . f\,(v_1\,q\,p \oplus p'))\,(v_2\,q\,p) \tag{mul-$S$-$s_3$}\\
    v_1\,q\,(p \oplus p') &= v_1\,q\,p \oplus v_1\,(v_2\,q\,p)\, p' \tag{mul-$S$-$p_1$}\\
    v_2\,q\,(p \oplus p') &= v_2\,(v_2\,q\,p)\,p' \tag{mul-$S$-$p_2$}\\
    \\
    u_2\, s\, (\lambda \_. \iota^T) &= \lconst{s} \tag{unit-$\iota T$-$s_2$}\\
    v_1\, \imparg{s}\, \imparg{\lambda \_. \iota^T}\, q\, p &= p \tag{unit-$\iota T$-$p_1$}\\
    v_2\, \imparg{s}\, \imparg{\lambda \_. \iota^T}\, q\, p &= q \tag{unit-$\iota T$-$p_2$}\\
    u_2\, s\, (\lambda p . \sigma^T\, (f\, p)\, (g \circ (p,-)))\, q &= u_2\, (u_2\, s\, f\, (\pr^T_1\, q))\, (g \circ v\, q)\, (\pr^T_2\, q) \tag{mul-$T$-$s_2$}\\
    v_1\, (\pr^T_1\, q)\, (v_1\, (\pr^T_2\, q)\, p) &= v_1\, q\, p \tag{mul-$T$-$p_1$}\\
    v_2\, (\pr^T_1\, q)\, (v_1\, (\pr^T_2\, q)\, p) &= \pr^T_1\, (v_2\, q\, p) \tag{mul-$T$-$p_{21}$}\\
    v_2\, (\pr^T_2\, q)\, p &= \pr^T_2\, (v_2\, q\, p) \tag{mul-$T$-$p_{22}$}
\end{align*}
\end{definition}

Notice that the first six equations are taken from directed container distributive laws, and the remaining equations from monadic container distributive laws. As in directed container distributive laws, the unit-$oS$-$s$ equality fully determines $u_1$.

Directed-monadic container distributive laws (of a directed container over a monadic container) can be derived in a similar fashion.

\begin{definition}[\flink{DirectedMndDistributiveLaw.html\#360}] \label{cm-cont-distr-law} 
Let $(\cont{S}{P}, \iota, \sigma, \pr)$ be a monadic container and $(\cont{T}{Q}, o , \oplus, \downarrow)$ be a directed container. We define a \ul{directed-monadic container distributive law} of $\cont{T}{Q}$ over $\cont{S}{P}$ as the data
\begin{align*}
    &u_1 : \nprod_{s : S} (P\, s \to T) \to T \\
    &u_2 : \nprod_{s : S} \nprod_{f : P\, s \to T} Q\, (u_1\, s\, f) \to S \\
    &v_1 : \nprod_{\imparg{s : S}} \nprod_{\imparg{f : P\, s \to T}} \nprod_{q : Q\, (u_1\, s\, f)} P\, (u_2\, s\, f\, q) \to P\, s \\
    &v_2 : \nprod_{\imparg{s : S}} \nprod_{\imparg{f : P\, s \to T}} \nprod_{q : Q\, (u_1\, s\, f)} \nprod_{p : P\, (u_2\, s\, f\, q)} Q\, (f\, (v_1\, q\, p))
\end{align*}
which satisfy the following equalities.
\begin{align*}
    u_1\, \iota^S\, (\lconst{t}) &= t \tag{unit-$\iota S$-$s_1$}\\
    u_2\, \iota^S\, (\lconst{t}) &=\lconst{\iota^S} \tag{unit-$\iota S$-$s_2$}\\
    v_1\, \imparg{\iota^S}\, \imparg{\lambda \_. t}\, q\, p &= p \tag{unit-$\iota S$-$p_1$}\\
    v_2\, \imparg{\iota^S}\, \imparg{\lambda \_. t}\, q\, p &= q \tag{unit-$\iota S$-$p_2$}\\
    u_1\, (\sigma^S\, s\, f)\, (g \circ \pr^S) &= u_1\, s\, (\lambda p . u_1\, (f\, p)\, (g \circ (p,-))) \tag{mul-$S$-$s_1$}\\
    u_2\, (\sigma^S\, s\, f)\, (g \circ \pr^S)\, q &= \sigma^S\,(u_2\, s\, (\lambda p .u_1\, (f\, p)\, (g \circ (p,-)))\, q)\\ 
    &\qquad\;\;\, (\lambda p . u_2\, (f\, (v_1\, q\, p))\, (g \circ (v_1\, q\, p , -))\, (v_2\, q\, p)) \tag{mul-$S$-$s_2$}\\  
    \pr^S_1\, (v_1\, q\, p) &= v_1\, q\, (\pr^S_1\, p) \tag{mul-$S$-$p_1$}\\
    \pr^S_2\, (v_1\, q\, p) &= v_1\, (v_2\, q\, (\pr^S_1\, p))\, (\pr^S_2\, p) \tag{mul-$S$-$p_{21}$}\\
    v_2\, q\, p &= v_2\, (v_2\, q\, (\pr^S_1\, p))\, (\pr^S_2\, p) \tag{mul-$S$-$p_{22}$}\\
    \\
    u_2\,s\,f\,(o\,\imparg{u_1\,s\,f}) &= s \tag{unit-$oT$-$s$}\\
    v_1\,(o\,\imparg{u_1\,s\,f})\,p &= p \tag{unit-$oT$-$p_1$}\\
    v_2\,(o\,\imparg{u_1\,s\,f})\,p &= o\,\imparg{s} \tag{unit-$oT$-$p_2$}\\
    u_1\,s\,f \downarrow q &= u_1\,(u_2\,s\,f\,q)\, (\lambda p . f\,(v_1\,q\,p) \downarrow v_2\,q\,p) \tag{mul-$T$-$s_1$}\\
    u_2\,s\,f\,(q \oplus q') &= u_2\,(u_2\,s\,f\,q)\,(\lambda p . f\,(v_1\,q\,p) \downarrow v_2\,q\,p)\,q' \tag{mul-$T$-$s_2$}\\
    v_1\,(q \oplus q')\,p &= v_1\,q\,(v_1\,q'\,p) \tag{mul-$T$-$p_1$}\\
    v_2\,(q \oplus q')\,p &= v_2\,q\,(v_1\,q'\,p) \oplus v_2\,q'\,p \tag{mul-$T$-$p_2$}
\end{align*}
\end{definition}

\begin{example}[\flink{DistributiveLawExamples.html\#14080}] \label{writer-reader-mixed-law} 
	There is a monadic-directed distributive law of the reader monadic container $\cont{\top}{\lconst{B}}$ over the writer directed container $\cont{A}{\lconst{\top}}$ for any sets $A$ and $B$, given by
\begin{align*}
    &u\,a\,f := (\star, \lconst{a})\\
    &v\,b\,\star := (\star, b).
\end{align*}
\end{example}

\begin{lemma}[\flink{DistributiveLawExamples.html\#15411}] \label{writer-reader-mixed-law-unique}  
The mixed distributive law in \cref{writer-reader-mixed-law} is the unique one of  $\cont{\top}{\lconst{B}}$ over $\cont{A}{\lconst{\top}}$.
\begin{proof}
    Follows from the isomorphism $(\top \to \top) \cong \top$ and then directly from the unit-$\iota T$ equations of monadic-directed distributive laws.
\end{proof}
\end{lemma}

It is thematically appropriate to check if these mixed distributive laws correspond to any known constructions on monoids. 

For monadic-directed distributive laws, we can do this by specialising the mixed distributive law to the case of $B \lhd \lconst{\top}$ over $\top \lhd \lconst{A}$, where $\top \lhd \lconst{A}$ is the reader directed container for a monoid $(A , e^A , \oplus^A)$, and $B \lhd \lconst{\top}$ is the writer monadic container for a monoid $(B , e^B , \oplus^B)$. $u_2$ and $v_2$ trivialise, $u_1$ is uniquely defined as $u_1\,s\,f = f\, e^A$, and the only parameter we are left with is \[v_1 : \nprod_{\imparg{\star : \top}} \nprod_{\imparg{f : A \to B}} \nprod_{\star : \top}\, \nprod_{a : A}\, A.\]
Possible values of $v_1$ in this case are in bijection with functions $\alpha : (A \to B) \to A \to A$, satisfying the equations:
\begin{align*}
    \alpha\,f\,e^A &= e^A\\
    \alpha\,f\,(a \oplus^A a') &= \alpha\,f\,a \oplus^A \alpha\,(\lambda x . f\,(\alpha\,f\,a \oplus^A x))\, a'\\
    \alpha\,(\lconst{e^B})\,a &= a\\
    \alpha\,(\lambda x . f\,x \oplus^B g\,x)\, a &= \alpha\,f\,(\alpha\,(\lambda x . g\,(\alpha\,f\,x))\, a).
\end{align*}

Counterintuitively, we do not end up with a matching pair of monoid actions! Instead we have something that we could call a `functional monoid action'\ (\flink{DistributiveLawExamples.html\#16619})  of the function space $A \to B$ on the monoid $A$. We have not come across this construction in the literature, but we would not be surprised if it was already known to algebraists.

Despite this, when we specialise directed-monadic container distributive laws to those between writer monadic containers and reader directed containers, we find that they \emph{are} in bijection with matching pairs of monoid actions. 

In summary, by restricting distributive laws to those between certain monadic and directed containers (those that are in bijection with monoids), we obtain corresponding constructions on monoids. The table below records the constructions obtained by considering distributive laws of each row container over each column container.
\\

\begin{tabular}{|c|c|c|}
\hline
 & Writer monadic container & Reader directed container \\
\hline
Writer monadic container & Matching pairs & Functional monoid actions  \\
\hline
Reader directed container & Matching pairs & Matching pairs \\
\hline
\end{tabular}
\\

This highlights an interesting asymmetry, however the reason for this asymmetry is not immediately obvious to us.

\section{A no-go theorem}\label{sec:no-go}

To show that our characterisation in \cref{m-cont-distr-law} is amenable to developing theorems concerning non-existence of distributive laws, we look to Zwart and Marsden \cite{zwart-no-go-theorems} for approaches to developing no-go theorems that we may be able to emulate.

To do this, we first need to translate properties of and statements about algebraic theories, into those for monadic containers. Our strategy for this translation was to \emph{very loosely} view aspects of a monadic container as analogous to aspects of an algebraic theory: 
\begin{itemize}
    \item shapes are `terms'
    \item positions are `variables within a term'
    \item $\sigma$ is the `substitution operator'
    \item $\iota$ is a `variable placeholder'
    \item $\pr$ is a map from `variable positions in a term resulting from a substitution' to `variable positions in the terms that were involved in that substitution'.
\end{itemize}

The point of this is not to form a rigorous connection between monadic containers and algebraic theories, but to see if there are common patterns that can be quickly taken advantage of. As it turns out, this loose perspective provides sufficient intuition to emulate the ``too many constants'' theorem about non-existence of certain composite algebraic theories in \cite[Theorem 4.6]{zwart-no-go-theorems} in our setting.

\begin{definition}[\flink{NoGoTheorem.html\#698}]
    Given a monadic container $(\cont{S}{P}, \iota, \sigma, \pr)$ we call any shape $s : S$ where
    $P\,s \cong \bot$ a \ul{constant shape}.
\end{definition}

\begin{definition}[\flink{NoGoTheorem.html\#870}] \label{s3-property}
    A monadic container $(\cont{S}{P}, \iota, \sigma, \pr)$ satisfies the \ul{(S3) property} if there exist $s : S$ and $f : P\,s \to S$ such that $P\,s$ is non-empty, equality on $P\,s$ is decidable, and for any $p : P\,s$
    \[
    \sigma\,s\,f^p = \iota
    \]
    and for any $p' : P\,(\sigma\,s\,f^p)$ we have
    \[
    \pr_1\,\imparg{s}\,\imparg{f^p}\,p' = p
    \]
    where we use the notation 
    \[
    f^p := \lambda p' . \begin{cases}
        \iota & \text{if } p = p'\\
        f\, p' & \text{otherwise}
    \end{cases}
    \]
\end{definition}

The name of this property is taken from the analogous property of algebraic theories in \cite[Section 4]{zwart-no-go-theorems}. In particular, the list monadic container $\cont{\nat}{\Fin}$ satisfies (S3) by taking any $n : \nat$ except $0$, and taking $f := \lconst{0}$. Since you can pick any $n : \nat$ where $\Fin\,n$ is non-empty, this container actually satisfies a stronger property, directly analogous to (S4) in \cite[Section 4]{zwart-no-go-theorems}. An interesting almost-example is the state monadic container $\cont{(S \to S)}{\lconst{S}}$ for $S$ with decidable equality, which satisfies the shape equation of (S3) by taking $s := \lambda x . x$ and $f := \lambda x y . y$, but \emph{not} the position equation (unless $S$ is trivial).

\begin{lemma}[Composite (S3), \flink{NoGoTheorem.html\#1430}] \label{composite-s3}
Let $(\cont{S}{P}, \iota^S, \sigma^S, \pr^S)$ be a monadic container satisfying (S3) for some $s : S$ and $f : P\,s \to S$, and $(\cont{T}{Q}, \iota^T, \sigma^T, \pr^T)$ be a monadic container. Assume we have a monadic container distributive law $(u_1, u_2, v_1, v_2)$ of $\cont{S}{P}$ over $\cont{T}{Q}$. Then, for any $t: T$ and $p : P\,s$,
\[
u_1\,s\,\Big (\lambda p' . \begin{cases}
    t & \text{if } p' = p\\
    \iota^T & \text{otherwise}
\end{cases}\Big ) = t.
\]
\begin{proof} 
    Proof included in \cref{appendixC}.
\end{proof}
\end{lemma}

\begin{theorem}[Too many constants, \flink{NoGoTheorem.html\#7534}] \label{mult-zeros}
Let $(\cont{S}{P}, \iota^S, \sigma^S, \pr^S)$ be a monadic container satisfying (S3) for some $s : S$ and $f : P\,s \to S$, such that we have two distinct positions $p, p' : P\,s$, and let $(\cont{T}{Q}, \iota^T, \sigma^T, \pr^T)$ be a monadic container. Assume we have a monadic container distributive law $(u_1, u_2, v_1, v_2)$ of $\cont{S}{P}$ over $\cont{T}{Q}$. Then $\cont{T}{Q}$ cannot have more than one distinct constant shape.
\begin{proof} 
    Proof included in \cref{appendixC}.
\end{proof}
\end{theorem}

With this theorem we can obtain the known result that there are no distributive laws of the list monad over the coproduct monad.

\begin{example}
Let $E$ be a set with at least two elements $e_1,e_2 : E$ such that $e_1 \neq e_2$. By \cref{mult-zeros}, there is no monadic container distributive law of the list monadic container $\cont{\nat}{\Fin{}}$\hspace{-2.5pt} over the coproduct monadic container $\cont{(\top + E)}{\Tr}$.
\end{example}

The scope of this theorem is slightly different to Zwart and Marsden's theorem of the same name, as it concerns a different class of monads -- those presentable as monadic containers, rather than algebraic theories. 
It is possible to represent any monad on $\cat{\Set}$ as a \emph{generalised} algebraic theory \cite{manes-algebraic-theories}, but it is not immediately clear how to transfer no-go theorems for compositions of algebraic theories into that setting.

\section{Conclusion}\label{sec:conclusion}

We have characterised monadic, monadic-directed, and directed-monadic container distributive laws, and have motivated their use for development of uniqueness and existence proofs. Further, we have shown such proofs to be amenable to mechanisation in a proof assistant.

Looking at the constructions on monoids that each of these distributive laws generalise, we note a curious asymmetry. Monadic-directed container distributive laws correspond to the notion of `functional monoid action', as opposed to matching pairs of monoid actions, which all other kinds of distributive law we consider correspond to.

Our work dualises that of Ahman and Uustalu in \cite{ahman-distr-laws}, completing the set of characterisations for container (co)monads and their distributive laws, but there are clearly further directions to explore. All of our characterisations and those in \cite{ahman-directed-cont,ahman-distr-laws,uustalu-monad-cont} could be extended to symmetric (or groupoid) containers \cite{gylterud-symm-cont}, or further to categorified containers \cite{gylterud-symm-cont,altenkirch-hott-uf-abstract} to describe a larger class of (co)monads and their distributive laws. For example, it seems possible to represent the finite multiset monad using groupoid containers. We intend to explore this direction in future work.

In \cref{predicate-univ}, we touch on the connection between type universes and monadic containers. This perspective could be further explored, using this work as a starting point. For example, one could consider developing a notion of composition for type universes with not only codes for $\top$ and $\Sigma$, but also for $\Pi$ types.

\bibliography{refs}

\newpage
\appendix

\crefalias{section}{appendix}

\section{Appendix}\label{appendixB}

\allowdisplaybreaks
\begin{proof}[Associativity proofs of \cref{distr-law-to-composite}]\let\qed\relax
Almost all parts of \cref{distr-law-to-composite} have been formalised and can be found here: \flink{MndDistrLawToCompatibleComposite.html\#8951}, aside from the associativity equalities (for technical reasons), which we prove below instead. 
Sub-expressions with the same colour were rewritten by applying the same position equality.
\begin{align*}
    &\zigma{}{(a_1 , a_2)}{(\lambda p . \zigma{}{(b_1\ p , b_2\ p)}{\langle c_1\ p , c_2\ p \rangle})}\\
    \coloneqq &(
    \zigma{S}{a_1}{(\lambda y . u_1\ (a_2\ y)\ (\lambda z . \zigma{S}{(b_1\ (y , z))}{(\lambda y' . u_1\ (b_2\ (y , z)\ y')\ (\lambda z' . c_1\ (y , z)\ (y' , z')))}))} ,\\
    &\lambda y . \sigma^T\\
        &\qquad (u_2\ (a_2\ (\pr_1^S\ y))\\ 
            &\qquad\qquad (\lambda z . \zigma{S}{(b_1\ (\pr_1^S\ y , z))}{(\lambda y' . u_1\ (b_2\ (\pr_1^S\ y , z)\ y')\ (\lambda z' . c_1\ (\pr_1^S\ y , z)\ (y' , z')))})\\
            &\qquad\qquad (\pr_2^S\ y))\\
        &\qquad (\lambda z . \sigma^T\\
            &\qquad\qquad (u_2\ (b_2\ (\pr_1^S\ y , v_1\ (\pr_2^S\ y)\ z)\ (\pr_1^S\ (v_2\ (\pr_2^S\ y)\ z)))\\
                &\qquad\qquad\qquad (\lambda z' . c_1\ (\pr_1^S\ y , v_1\ (\pr_2^S\ y)\ z)\ (\pr_1^S\ (v_2\ (\pr_2^S\ y)\ z) , z'))\\
                &\qquad\qquad\qquad (\pr_2^S\ (v_2\ (\pr_2^S\ y)\ z)))\\
            &\qquad\qquad (\lambda z' . c_2\ (\pr_1^S\ y , v_1\ (\pr_2^S\ y)\ z)\\
            &\qquad\qquad\qquad (\pr_1^S\ (v_2\ (\pr_2^S\ y)\ z) , v_1\ (\pr_2^S\ (v_2\ (\pr_2^S\ y)\ z))\ z')\\
            &\qquad\qquad\qquad (v_2\ (\pr_2^S\ (v_2\ (\pr_2^S\ y)\ z))\ z'))))\\
    =&\text{[$\sigma^T$ and $\pr^T$ associativity]}\\
    &(\zigma{S}{a_1}{(\lambda y . u_1\ (a_2\ y)\ (\lambda z . \zigma{S}{(b_1\ (y , z))}{(\lambda y' . u_1\ (b_2\ (y , z)\ y')\ (\lambda z' . c_1\ (y , z)\ (y' , z')))}))} ,\\
    &\lambda y . \sigma^T\\
        &\qquad (\sigma^T\\
        &\qquad\quad (u_2\ (a_2\ (\pr_1^S\ y))\\ 
            &\qquad\quad\qquad (\lambda z . \zigma{S}{(b_1\ (\pr_1^S\ y , z))}{(\lambda y' . u_1\ (b_2\ (\pr_1^S\ y , z)\ y')\ (\lambda z' . c_1\ (\pr_1^S\ y , z)\ (y' , z')))})\\
            &\qquad\quad\qquad (\pr_2^S\ y))\\
        &\qquad\quad (\lambda z . u_2\ (b_2\ (\pr_1^S\ y , v_1\ (\pr_2^S\ y)\ z)\ (\pr_1^S\ (v_2\ (\pr_2^S\ y)\ z)))\\
        &\qquad\qquad\qquad (\lambda z' . c_1\ (\pr_1^S\ y , v_1\ (\pr_2^S\ y)\ z)\ (\pr_1^S\ (v_2\ (\pr_2^S\ y)\ z) , z'))\\
        &\qquad\qquad\qquad (\pr_2^S\ (v_2\ (\pr_2^S\ y)\ z)))\\
        &\qquad\quad (\lambda z . c_2\ (\pr_1^S\ y , v_1\ (\pr_2^S\ y)\ (\hlred{\pr_1^T\ z}))\\
        &\qquad\qquad\qquad (\pr_1^S\ (v_2\ (\pr_2^S\ y)\ (\hlred{\pr_1^T\ z})) , v_1\ (\pr_2^S\ (v_2\ (\pr_2^S\ y)\ (\hlred{\pr_1^T\ z})))\ (\hlgreen{\pr_2^T\ z}))\\
        &\qquad\qquad\qquad (v_2\ (\pr_2^S\ (v_2\ (\pr_2^S\ y)\ (\hlred{\pr_1^T\ z})))\ (\hlgreen{\pr_2^T\ z}))))\\
    =&\text{[mul-$T$ equations]}\\
    &(\sigma^S\ a_1\ (\lambda y . \sigma^S\\
    &\qquad\qquad (u_1\ (a_2\ y)\ (\lambda z . b_1\ (y , z)))\\
    &\qquad\qquad (\lambda y' . u_1\\
    &\qquad\qquad\qquad (u_2\ (a_2\ y)\ (\lambda z . b_1\ (y , z))\ y')\\
    &\qquad\qquad\qquad (\lambda z. u_1\ (b_2\ (y , v_1\ y'\ z)\ (v_2\ y'\ z))\ (\lambda z' . c_1\ (y , v_1\ y'\ z)\ (v_2\ y'\ z , z'))))),\\
    &\lambda y . \sigma^T\\
        &\qquad (\sigma^T\\ 
        &\qquad\qquad (u_2\ (u_2\ (a_2\ (\pr_1^S\ y))\ (\lambda z . b_1\ (\pr_1^S\ y , z))\ (\pr_1^S\ (\pr_2^S\ y)))\\
        &\qquad\qquad\qquad (\lambda z. u_1\ (b_2\ (\pr_1^S\ y , v_1\ (\pr_1^S\ (\pr_2^S\ y))\ z)\ (v_2\ (\pr_1^S\ (\pr_2^S\ y))\ z))\\
        &\qquad\qquad\qquad\qquad\quad\; (\lambda z' . c_1\ (\pr_1^S\ y , v_1\ (\pr_1^S\ (\pr_2^S\ y))\ z)\ (v_2\ (\pr_1^S\ (\pr_2^S\ y))\ z , z')))\\
        &\qquad\qquad\qquad (\pr_2^S\ (\pr_2^S\ y)))\\
        &\qquad\qquad (\lambda z . u_2\ (b_2\ (\pr_1^S\ y , \hlred{v_1\ (\pr_1^S\ (\pr_2^S\ y))\ (v_1\ (\pr_2^S\ (\pr_2^S\ y))\ z)})\\ 
        &\qquad\qquad\qquad\qquad\quad (\hlgreen{v_2\ (\pr_1^S\ (\pr_2^S\ y))\ (v_1\ (\pr_2^S\ (\pr_2^S\ y))\ z)}))\\
        &\qquad\qquad\qquad\quad\; (\lambda z' . c_1 (\pr_1^S\ y , \hlred{v_1\ (\pr_1^S\ (\pr_2^S\ y))\ (v_1\ (\pr_2^S\ (\pr_2^S\ y))\ z)})\\
        &\qquad\qquad\qquad\qquad\qquad\; (\hlgreen{v_2\ (\pr_1^S\ (\pr_2^S\ y))\ (v_1\ (\pr_2^S\ (\pr_2^S\ y))\ z)} , z'))\\
        &\qquad\qquad\qquad\quad\; (\hlblue{v_2\ (\pr_2^S\ (\pr_2^S\ y))\ z})))\\
        &\qquad (\lambda z . c_2\ (\pr_1^S\ y , \hlred{v_1\ (\pr_1^S\ (\pr_2^S\ y))\ (v_1\ (\pr_2^S\ (\pr_2^S\ y))\ (\pr_1^T\ z))})\\
        &\qquad\qquad\quad\, (\hlgreen{v_2\ (\pr_1^S\ (\pr_2^S\ y))\ (v_1\ (\pr_2^S\ (\pr_2^S\ y))\ (\pr_1^T\ z))}, \\
        &\qquad\qquad\quad\;\; v_1\ (\hlblue{v_2\ (\pr_2^S\ (\pr_2^S\ y))\ (\pr_1^T\ z)})\ (\pr_2^T\ z))\\
        &\qquad\qquad\quad\, (v_2\ (\hlblue{v_2\ (\pr_2^S\ (\pr_2^S\ y))\ (\pr_1^T\ z)})\ (\pr_2^T\ z))))\\
    =&\text{[mul-$S$ equations]}\\
    &(\sigma^S\ a_1\ (\lambda y . \sigma^S\\
    &\qquad\qquad\qquad (u_1\ (a_2\ y)\ (\lambda z . b_1\ (y , z)))\\
    &\qquad\qquad\qquad (\lambda y' . u_1\ (\sigma^T\ (u_2\ (a_2\ y)\ (\lambda z . b_1\ (y , z))\ y')\ (\lambda z . b_2\ (y , v_1\ y'\ z)\ (v_2\ y'\ z)))\ \\
    &\qquad\qquad\qquad\qquad\quad\;\; (\lambda z. c_1\ (y , v_1\ y'\ (\pr_1^T\ z))\ (v_2\ y'\ (\pr_1^T\ z) , \pr_2^T\ z)))), \\
    &\lambda y . \sigma^T\\
        &\qquad (u_2\ (\sigma^T\ (u_2\ (a_2\ (\pr_1^S\ y))\ (\lambda z . b_1\ (\pr_1^S\ y , z))\ (\pr_1^S\ (\pr_2^S\ y)))\\
        &\qquad\qquad\quad\;\;\; (\lambda z . b_2\ (\pr_1^S\ y , v_1\ (\pr_1^S\ (\pr_2^S\ y))\ z)\ (v_2\ (\pr_1^S\ (\pr_2^S\ y))\ z)))\\
        &\qquad\qquad (\lambda z. c_1\ (\pr_1^S\ y , v_1\ (\pr_1^S\ (\pr_2^S\ y))\ (\pr_1^T\ z))\ (v_2\ (\pr_1^S\ (\pr_2^S\ y))\ (\pr_1^T\ z) , \pr_2^T\ z))\\
        &\qquad\qquad (\pr_2^S\ (\pr_2^S\ y)))\\
        &\qquad (\lambda z . c_2\ (\pr_1^S\ y , v_1\ (\pr_1^S\ (\pr_2^S\ y))\ (\hlred{\pr_1^T\ (v_1\ (\pr_2^S\ (\pr_2^S\ y))\ z)}))\\
        &\qquad\qquad\quad (v_2\ (\pr_1^S\ (\pr_2^S\ y))\ (\hlred{\pr_1^T\ (v_1\ (\pr_2^S\ (\pr_2^S\ y))\ z)}) , \hlgreen{\pr_2^T\ (v_1\ (\pr_2^S\ (\pr_2^S\ y))\ z)})\\
        &\qquad\qquad\quad (\hlblue{v_2\ (\pr_2^S\ (\pr_2^S\ y))\ z})))\\
    =&\text{[$\sigma^S$ and $\pr^S$ associativity equations]}\\
    &(\sigma^S\ (\sigma^S\ a_1\ (\lambda y . u_1\ (a_2\ y)\ (\lambda z . b_1\ (y , z))))\\ 
    &\qquad (\lambda y . u_1\ (\sigma^T\ (u_2\ (a_2\ (\pr_1^S\ y))\ (\lambda z . b_1\ (\pr_1^S\ y , z))\ (\pr_2^S\ y))\\
    &\qquad\qquad\qquad\quad\, (\lambda z . b_2\ (\pr_1^S\ y , v_1\ (\pr_2^S\ y)\ z)\ (v_2\ (\pr_2^S\ y)\ z)))\\
    &\qquad\qquad\quad\; (\lambda z. c_1\ (\pr_1^S\ y , v_1\ (\pr_2^S\ y)\ (\pr_1^T\ z))\ (v_2\ (\pr_2^S\ y)\ (\pr_1^T\ z) , \pr_2^T\ z))) ,\\
    &\lambda y . \sigma^T\ (u_2\ (\sigma^T\ (u_2\ (a_2\ (\hlred{\pr_1^S\ (\pr_1^S\ y)}))\ (\lambda z . b_1\ (\hlred{\pr_1^S\ (\pr_1^S\ y)} , z))\ (\hlgreen{\pr_2^S\ (\pr_1^S\ y)}))\\
        &\qquad\qquad\qquad\;\;\; (\lambda z . b_2\ (\hlred{\pr_1^S\ (\pr_1^S\ y)} , v_1\ (\hlgreen{\pr_2^S\ (\pr_1^S\ y)})\ z)\ (v_2\ (\hlgreen{\pr_2^S\ (\pr_1^S\ y)})\ z)))\\
        &\qquad\qquad\;\;\; (\lambda z. c_1\ (\hlred{\pr_1^S\ (\pr_1^S\ y)} , v_1\ (\hlgreen{\pr_2^S\ (\pr_1^S\ y)})\ (\pr_1^T\ z))\\ 
        &\qquad\qquad\qquad\qquad (v_2\ (\hlgreen{\pr_2^S\ (\pr_1^S\ y)})\ (\pr_1^T\ z) , \pr_2^T\ z))\ (\hlblue{\pr_2^S\ y}))\\
        &\qquad\quad (\lambda z . c_2\ (\hlred{\pr_1^S\ (\pr_1^S\ y)} , v_1\ (\hlgreen{\pr_2^S\ (\pr_1^S\ y)})\ (\pr_1^T\ (v_1\ (\hlblue{\pr_2^S\ y})\ z)))\\
        &\qquad\qquad\qquad (v_2\ (\hlgreen{\pr_2^S\ (\pr_1^S\ y)})\ (\pr_1^T\ (v_1\ (\hlblue{\pr_2^S\ y})\ z)) , \pr_2^T\ (v_1\ (\hlblue{\pr_2^S\ y})\ z))\\
        &\qquad\qquad\qquad (v_2\ (\hlblue{\pr_2^S\ y})\ z)))\\
    &=: \zigma{}{(\sigma\ (a_1 , a_2)\ (\lambda p .(b_1\ p , b_2\ p)))}{(\lambda p . (c_1\ (\pr_1\ p)\ (\pr_2\ p) , c_2\ (\pr_1\ p)\ (\pr_2\ p)))}
\end{align*}
\begin{align*}
    &\ \pr_1\ (\pr_1\ (p , q))\\
    \coloneqq&\ \pr_1^S\ (\pr_1^S\ p) , v_1\ (\pr_2^S\ (\pr_1^S\ p))\ (\pr_1^T\ (v_1\ (\pr_2^S\ p)\ (\pr_1^T\ q)))\\
    =&\ \pr_1^S\ p , v_1\ (\pr_1^S\ (\pr_2^S\ p))\ (\pr_1^T\ (v_1\ (\pr_2^S\ (\pr_2^S\ p))\ (\pr_1^T\ q)))
    &\tag*{[$\pr^S$ assoc.\ eqs.]}\\
    =&\ \pr_1^S\ p , v_1\ (\pr_1^S\ (\pr_2^S\ p))\ (v_1\ (\pr_2^S\ (\pr_2^S\ p))\ (\pr_1^T\ (\pr_1^T\ q)))
    &\tag*{[mul-$S$-p$_1$]}\\
    =&\ \pr_1^S\ p , v_1\ (\pr_2^S\ p)\ (\pr_1^T\ (\pr_1^T\ q))
    &\tag*{[mul-$T$-p$_1$]}\\
    =&\ \pr_1^S\ p , v_1\ (\pr_2^S\ p)\ (\pr_1^T\ q)
    &\tag*{[$\pr^T$ assoc.\ eqs.]}\\
    \eqqcolon&\ \pr_1\ (p , q)
\end{align*}
\begin{align*}
    &\ \pr_2\ (\pr_1\ (p , q))\\
    \coloneqq&\ v_2\ (\pr_2^S\ (\pr_1^S\ p))\ (\pr_1^T\ (v_1\ (\pr_2^S\ p)\ (\pr_1^T\ q))) , \pr_2^T\ (v_1\ (\pr_2^S\ p)\ (\pr_1^T\ q))\\
    =&\ v_2\ (\pr_1^S\ (\pr_2^S\ p))\ (\pr_1^T\ (v_1\ (\pr_2^S\ (\pr_2^S\ p))\ (\pr_1^T\ q))), \\
    & \qquad \pr_2^T\ (v_1\ (\pr_2^S\ (\pr_2^S\ p))\ (\pr_1^T\ q))  \tag*{[$\pr^S$ assoc.\ eqs.]}\\
    =&\ v_2\ (\pr_1^S\ (\pr_2^S\ p))\ (v_1\ (\pr_2^S\ (\pr_2^S\ p))\ (\pr_1^T\ (\pr_1^T\ q))) ,\\
    &\qquad\, v_1\ (v_2\ (\pr_2^S\ (\pr_2^S\ p))\ (\pr_1^T\ (\pr_1^T\ q)))\ (\pr_2^T\ (\pr_1^T\ q)) \tag*{[mul-$S$-p$_1$, mul-$S$-p$_{2 1}$]}\\
    =&\ \pr_1^S\ (v_2\ (\pr_2^S\ p)\ (\pr_1^T\ (\pr_1^T\ q))) ,\\
    &\qquad\;\; v_1\ (\pr_2^S\ (v_2\ (\pr_2^S\ p)\ (\pr_1^T\ (\pr_1^T\ q))))\ (\pr_2^T\ (\pr_1^T\ q)) \tag*{[mul-$T$-p$_{2 1}$, mul-$T$-p$_{2 2}$]}
    \\
    =&\ \pr_1^S\ (v_2\ (\pr_2^S\ p)\ (\pr_1^T\ q)) ,\\
    &\qquad\;\; v_1\ (\pr_2^S\ (v_2\ (\pr_2^S\ p)\ (\pr_1^T\ q)))\ (\pr_1^T\ (\pr_2^T\ q)) \tag*{[$\pr^T$ assoc.\ eqs.]}
    \\
    \eqqcolon&\ \pr_1\ (\pr_2\ (p , q))
\end{align*}
\begin{align*}
    &\ \pr_2\ (p , q)\\
    \coloneqq&\  v_2\ (\pr_2^S\ p)\ (\pr_1^T\ q) , \pr_2^T\ q\\
    =&\ v_2\ (\pr_2^S\ (\pr_2^S\ p))\ (\pr_1^T\ q) , \pr_2^T\ q
    &\tag*{[$\pr^S$ assoc.\ eqs.]}\\
    =&\ v_2\ (v_2\ (\pr_2^S\ (\pr_2^S\ p))\ (\pr_1^T\ (\pr_1^T\ q)))\ (\pr_2^T\ (\pr_1^T\ q)) , \pr_2^T\ q
    &\tag*{[mul-$S$-p$_{2 2}$]}\\
    =&\ v_2\ (\pr_2^S\ (v_2\ (\pr_2^S\ p)\ (\pr_1^T\ (\pr_1^T\ q))))\ (\pr_2^T\ (\pr_1^T\ q)) , \pr_2^T\ q
    &\tag*{[mul-$T$-p$_{2 2}$]}\\
    =&\ v_2\ (\pr_2^S\ (v_2\ (\pr_2^S\ p)\ (\pr_1^T\ q)))\ (\pr_1^T\ (\pr_2^T\ q)) , \pr_2^T\ (\pr_2^T\ q)
    &\tag*{[$\pr^T$ assoc.\ eqs.]}\\
    \eqqcolon&\  \pr_2\ (\pr_2\ (p , q)) 
    & \tag*{\textcolor{darkgray}{$\blacktriangleleft$}}\\
\end{align*}\vspace{-5pt}
\end{proof}

\section{Appendix}\label{appendixC}
The proofs in this appendix are for \cref{sec:no-go}.

\allowdisplaybreaks
\begin{proof}[Proof of \cref{composite-s3}]\let\qed\relax
	By the chain of equalities:
	\begin{align*}
		&u_1\,s\,\Big (\lambda p' . \begin{cases}
			t & \text{if } p' = p\\
			\iota^T & \text{otherwise}
		\end{cases}\Big )\\
		&= u_1\,s\,\Big (\lambda p' . \begin{cases}
			u_1\, \iota^S\, (\lconst{t}) & \text{if } p' = p\\
			u_1\, (f\,p')\, (\lconst{\iota^T}) & \text{otherwise}
		\end{cases}\Big )& \tag*{[\text{unit-$\iota S$-$s_1$ \& unit-$\iota T$-$s_1$}] \phantom{$\blacktriangleleft$}}\\
		&= u_1\,s\,\Big (\lambda p' . u_1\,\Big (\begin{cases}
			\iota^S & \text{if } p' = p\\
			f\,p' & \text{otherwise}
		\end{cases}\Big )
		\,\Big (\lconst{\begin{cases}
				t & \text{if } p' = p\\
				\iota^T & \text{otherwise}
		\end{cases}}\Big )\Big )\\
		&= u_1\,\Big (\sigma^S\,s\,\Big (\lambda p' . \begin{cases}
			\iota^S & \text{if } p' = p\\
			f\,p' & \text{otherwise}
		\end{cases}\Big )\Big )
		\,\Big (\lambda p'' . \begin{cases}
			t & \text{if } \pr^S_1\,p'' = p\\
			\iota^T & \text{otherwise}
		\end{cases}\Big )& \tag*{[\text{mul-$S$-$s_1$}] \phantom{$\blacktriangleleft$}}\\
		&= u_1\,\iota^S\,\Big (\lconst{\begin{cases}
				t & \text{if } p = p\\
				\iota^T & \text{otherwise}
		\end{cases}}\Big ) & \tag*{[\text{(S3)}]  \phantom{$\blacktriangleleft$}}\\
		& = t & \tag*{[\text{unit-$\iota S$-$s_1$}] \textcolor{darkgray}{$\blacktriangleleft$}}\\
	\end{align*}
\end{proof}

\begin{proof}[Proof of \cref{mult-zeros}]
	Assume we have two distinct constant shapes $t_0, t_1 : T$. We first have the chain of equalities:
	{\belowdisplayskip=2pt
	\begin{align*}
		&t_0\\
		&= \sigma^T\,t_0\,(\lconst{\iota^T}) & \tag*{[\cref{mcont}]}\\ 
		&= \sigma^T\,\Big (u_1\, s\,\Big (\lambda y . \begin{cases}
			t_0 & \text{if } y = p\\
			\iota^T & \text{otherwise}
		\end{cases}\Big )\Big )\\
		&\ \ \ \ \ \ \ \ \Big (\lambda y . u_1\,\Big (u_2\,s\,\Big (\lambda z . \begin{cases}
			t_0 & \text{if } z = p\\
			\iota^T & \text{otherwise}
		\end{cases}\Big )\,y\Big )
		\,\Big (\lambda z . \begin{cases}
			t_1 & \text{if } v_1\,y\,z = p'\\
			\iota^T & \text{otherwise}
		\end{cases}\Big )\Big ) & \tag*{[\cref{composite-s3}, $Q\,t_0 \to T$ is contractible and transporting over \cref{composite-s3} preserves this]}\\ 
		&= u_1\,s\,\Big (\lambda y . \sigma^T\,\Big (\begin{cases}
			t_0 & \text{if } y = p\\
			\iota^T & \text{otherwise}
		\end{cases}\Big )
		\,\Big (\lambda \_ . \begin{cases}
			t_1 & \text{if } y = p'\\
			\iota^T & \text{otherwise}
		\end{cases}\Big )\Big ) & \tag*{[\text{mul-$T$-$s_1$}]}\\
		&= u_1\,s\,\Bigg (\lambda y . \begin{cases}
			\sigma^T\, t_0\,(\lconst{\iota^T}) & \text{if } y = p\\
			\sigma^T\, \iota^T\,(\lconst{t_1}) & \text{if } y = p'\\
			\sigma^T\,\iota^T\,(\lconst{\iota^T}) & \text{otherwise}
		\end{cases}\Bigg ) & \tag*{[\text{$p \neq p'$ by assumption}]}\\
		&= u_1\,s\,\Bigg (\lambda y . \begin{cases}
			t_0 & \text{if } y = p\\
			t_1 & \text{if } y = p'\\
			\iota^T & \text{otherwise}
		\end{cases}\Bigg ) & \tag*{[\cref{mcont}]}\\  
	\end{align*}
	}
	Using the same steps, we can derive:
	\begin{align*}
		t_1 = u_1\,s\,\Bigg (\lambda y . \begin{cases}
			t_0 & \text{if } y = p\\
			t_1 & \text{if } y = p'\\
			\iota^T & \text{otherwise}
		\end{cases}\Bigg )
	\end{align*}
	and by composing these two equalities, we get that $t_0 = t_1$. This contradicts our assumption that $t_0$ and $t_1$ were distinct.
\end{proof}

Proof that the list monadic container in \cref{list-mcont} satisfies the (S3) property (\cref{s3-property}), for any $n : \mathbb{N}$ where $f = \lconst{0}$.
\begin{proof}
First notice that $\sigma\,n\,f^p = 1$ for any $p : \Fin\,n$. This means that the shape equality is satisfied, and also that we only have to consider $p' = 0$ for the position equality.

The sum $f^p\, 0 + \dots + f^p\, (i - 1)$ becomes $1$ exactly when $i = p$, which gives us that $\max\, \{ i \in [0..n)\ |\ f^p\, 0 + \dots + f^p\, (i - 1) \le 0 \} = p$.
Therefore, by definition, we have $\pr_1\,\imparg{n}\,\imparg{f^p}\,p' = p$.
\end{proof}

\end{document}